\documentclass[aps,prb,twocolumn,superscriptaddress,showpacs]{revtex4-1}
\usepackage[intlimits,sumlimits]{amsmath}
\usepackage{graphicx}
\usepackage{hyperref}

\providecommand\tsub[1]{\ensuremath{_{_{\text{#1}}}}} %
\providecommand\tP{\tsub{P}} %
\providecommand\tExt{\tsub{ext}} %
\providecommand\icm{\ensuremath{\text{cm}^{-1}}} % inverse centimeter
\providecommand\SiO[1]{SiO$_{#1}$} % Silicon Dioxide
\providecommand\eps[1]{\ensuremath{\epsilon_{#1}}} %
\providecommand\kz[1]{\ensuremath{k^{z}_{#1}}} %
\providecommand\bvec[1]{\ensuremath{\boldsymbol{#1}}} %
\providecommand\BUPhysics{Department of Physics, %
Boston University, %
590 Commonwealth Avenue, Boston, Massachusetts, 02215} %
\providecommand\UCSDPhysics{Department of Physics, %
University of California San Diego, %
9500 Gilman Drive, La Jolla, California 92093} %
\begin{document}
\title{Near-field spectroscopy of silicon dioxide thin films}

\author{L. M. Zhang}
\affiliation{\BUPhysics}

\author{G. O. Andreev}

\author{Z. Fei}
\author{A. S. McLeod}
\affiliation{\UCSDPhysics}

\author{G. Dominguez}
\author{M. Thiemens}
\affiliation{Department of Chemistry, University of California San Diego, 9500 Gilman Drive, La Jolla, California 92093}

\author{A. H. Castro~Neto}
\affiliation{Graphene Research Centre and Department of Physics, National University of Singapore, 2 Science Drive 3, 117542, Singapore}

\author{D. N. Basov}
\author{M. M. Fogler}
\affiliation{\UCSDPhysics}

\date{\today}

%Near-field scanning optical microscopy, 68.37.Uv
%Low-dimensional structures/phonons in, 63.22.-mP

\begin{abstract}
We analyze the results of scanning near-field infrared spectroscopy performed on thin films of a-\SiO2 on Si substrate.
The measured near-field signal exhibits surface-phonon resonances whose strength has a strong thickness dependence in the range from $2$ to $300\,\text{nm}$.
These observations are compared with calculations in which the tip of the near-field infrared spectrometer is modeled either as a point dipole or an elongated spheroid.
The latter model accounts for the antenna effect of the tip and gives a better agreement with the experiment.
Possible applications of the near-field technique for depth profiling of layered nanostructures are discussed.
\end{abstract}

\maketitle

%%%%%%%%%%%%%%%%%%%%%%%%%%%%%%%%%%%%%%%%%%%%%%%%%%%%%%%%%%%%%%%%%%%%%%%%%%%%%
\section{Introduction}
\label{sec:Introduction}
Scattering scanning near-field optical microscopy (s-SNOM)~\cite{Keilmann2004nfm, Novotny2006, Keilmann2009} is a powerful tool for probing local electromagnetic response of diverse materials. The s-SNOM achieves spatial resolution of $10$--$20\,\text{nm}$, which is especially valuable in the physically interesting infrared region~\cite{Basov2011eoc, Basov2011mfa} where the resolution of conventional spectroscopy is fundamentally limited by
a rather large wavelength $\lambda \sim 5$--$500\,\mu\text{m}$. The s-SNOM techniques have been rapidly advancing,~\cite{Amarie2011bia, Huth2011isn} which enabled their applications to imaging spectroscopy of complex oxides~\cite{Qazilbash2007mti, zhan2007tmi, Qazilbash2009Isa, Frenzel2009ies, Lai2010mpr, Qazilbash2011nio} and graphene.~\cite{Fei2011ino}

%%%%%%%%%%%%%%%%%%%%%%%%%%%%%%%%%%%%%%%%%%%%%%%%%%%%%%%%%%%%%%%%%%%%%%%%%%
% FIG. 1
%
\begin{figure}[b]
  \begin{center}
		\includegraphics[width=2.6in]{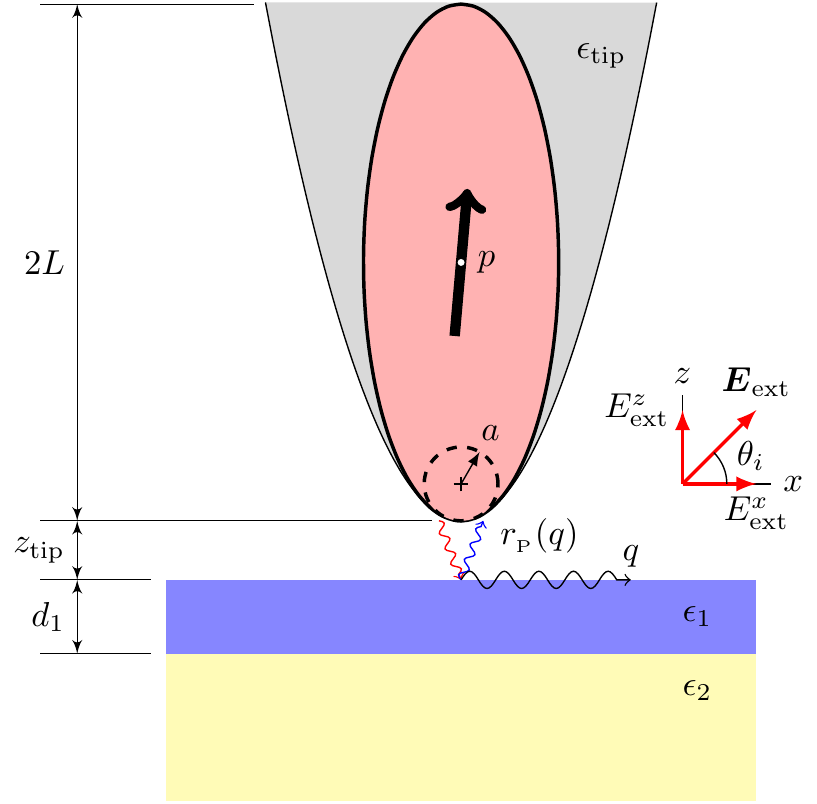}
  \end{center}
  \caption{(Color online) Schematics of an s-SNOM experiment.
  A scanned probe, modeled as a metallic spheroid with length $2 L$ and the apex curvature radius $a$, is positioned distance $z_{\text{tip}}$ above the sample.
  The sample contains a film of thickness $d_1$ and dielectric function $\eps1$, which is deposited on a bulk substrate with dielectric function $\eps2$.
  The system is illuminated by infrared field $\bvec{E}_{\text{ext}}$ at an angle of incidence $\theta$.
  Scattering of this radiation by the tip creates evanescent waves with large in-plane momenta $q \sim 1 / a$.
  The experiment measures the total radiating dipole $p$ of tip, which is determined by multiple reflections of the evanescent waves between the tip and sample.
  The reflections off the sample are characterized by the coefficient $r\tP(q, \omega)$.}
\label{fig:sSNOM}
\end{figure} %%%%%%%%%%%%%%%%%%%%%%%%%%%%%%%%%%%%%%%%%%%%%%%%%%%%%%%%%%%%%%%%%%%%%%%%%

The s-SNOM utilizes scattering of incident light by the 
tip of an atomic force microscope (AFM) positioned next to the probed sample (Fig.~\ref{fig:sSNOM}). The tip couples to the sample via evanescent waves of large in-plane momenta $q \sim 1 / a$, where $a$ is the tip radius of curvature (a few tens of nm). This is why the lateral resolution of the s-SNOM is determined primarily by $a$ rather than $\lambda$.~\cite{Lai2007afm, Huber2008TNF, Olmon2008nfi} 

One of the interesting open questions is the depth ($z$-coordinate) resolution of the s-SNOM probes. Previous experiments suggested that it is comparable to the lateral resolution $\sim a$, based on
imaging of small sub-surface particles.~\cite{Taubner2005nrs} Surprisingly, our recently near-field measurements of \SiO2 thin films have demonstrated that films as thick as several hundred nm have a response clearly different from that of the bulk material.~\cite{Andreev2011xxx} Thus, if instead of particles one has layers, then the s-SNOM is able to detect them at much larger depths.

In this paper these experimental results are re-analyzed and compared with two theoretical models, the conventional point-dipole approximation~\cite{Keilmann2004nfm, Hillendrand2002pel} and the spheroidal model. The former is very simple to implement but is also very crude. Predictably, it yields a bulk-like response of the s-SNOM signal as soon as the \SiO2 film thickness exceeds the tip radius, in disagreement with the experiment. A plausible reason for shortcomings of the point-dipole model is its failure to account for the strongly elongated shape of the tip. Such a tip acts as an optical antenna~\cite{Keilmann2004nfm, Novotny2006, Keilmann2009} that greatly enhances the electric field inside the tip-sample nanogap. Unfortunately, analytical models,~\cite{Cvitkovic2007amf, Moon2011qaa} that attempt to treat elongated tips do not apply to layered substrates. This compels us to study the problem numerically.

%%%%%%%%%%%%%%%%%%%%%%%%%%%%%%%%%%%%%%%%%%%%%%%%%%%%%%%%%%%%%%%%%%%%%%%%%%
% FIG. 2
%
\begin{figure*}
  %\begin{center}
    \includegraphics[width=2.3in]{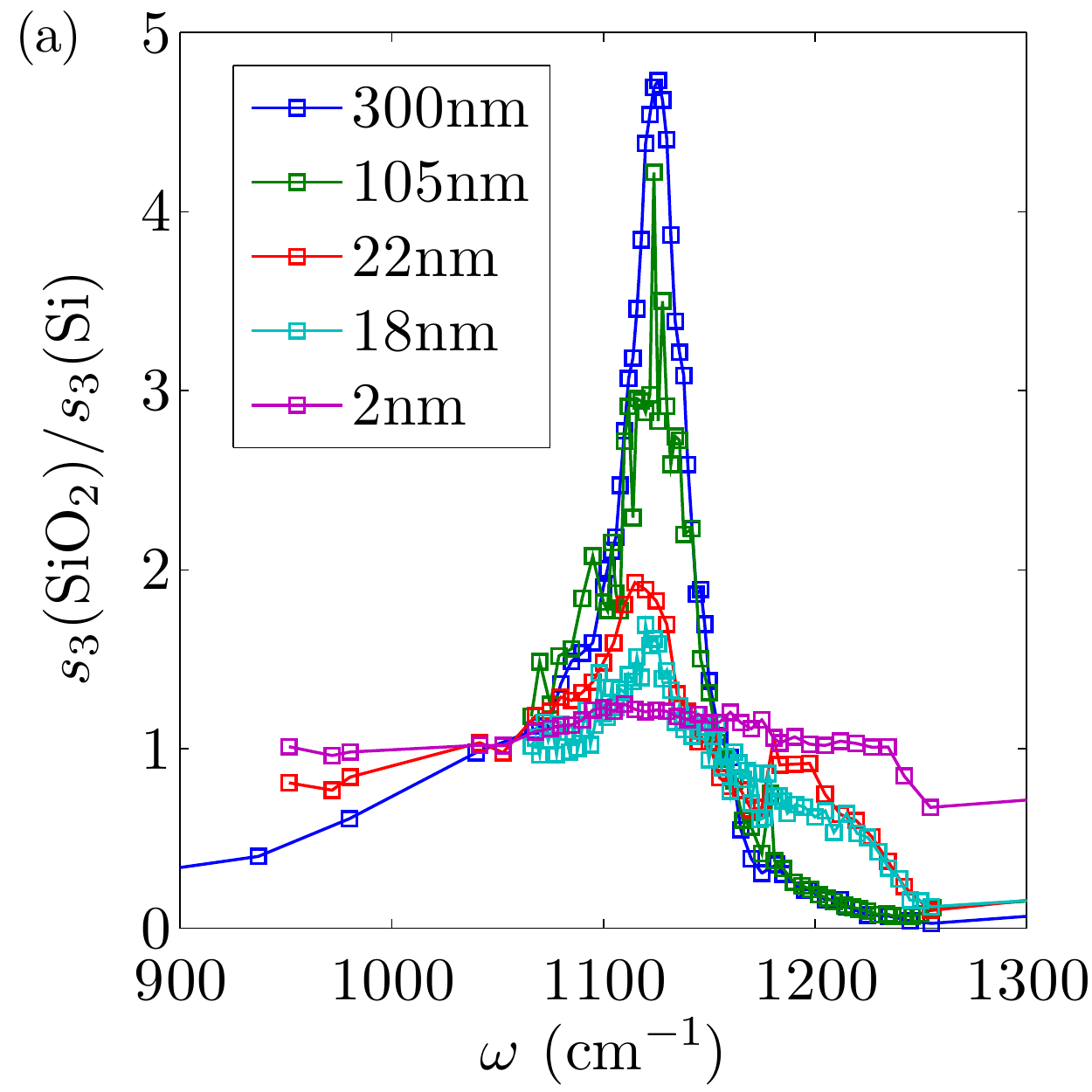}
    \includegraphics[width=2.3in]{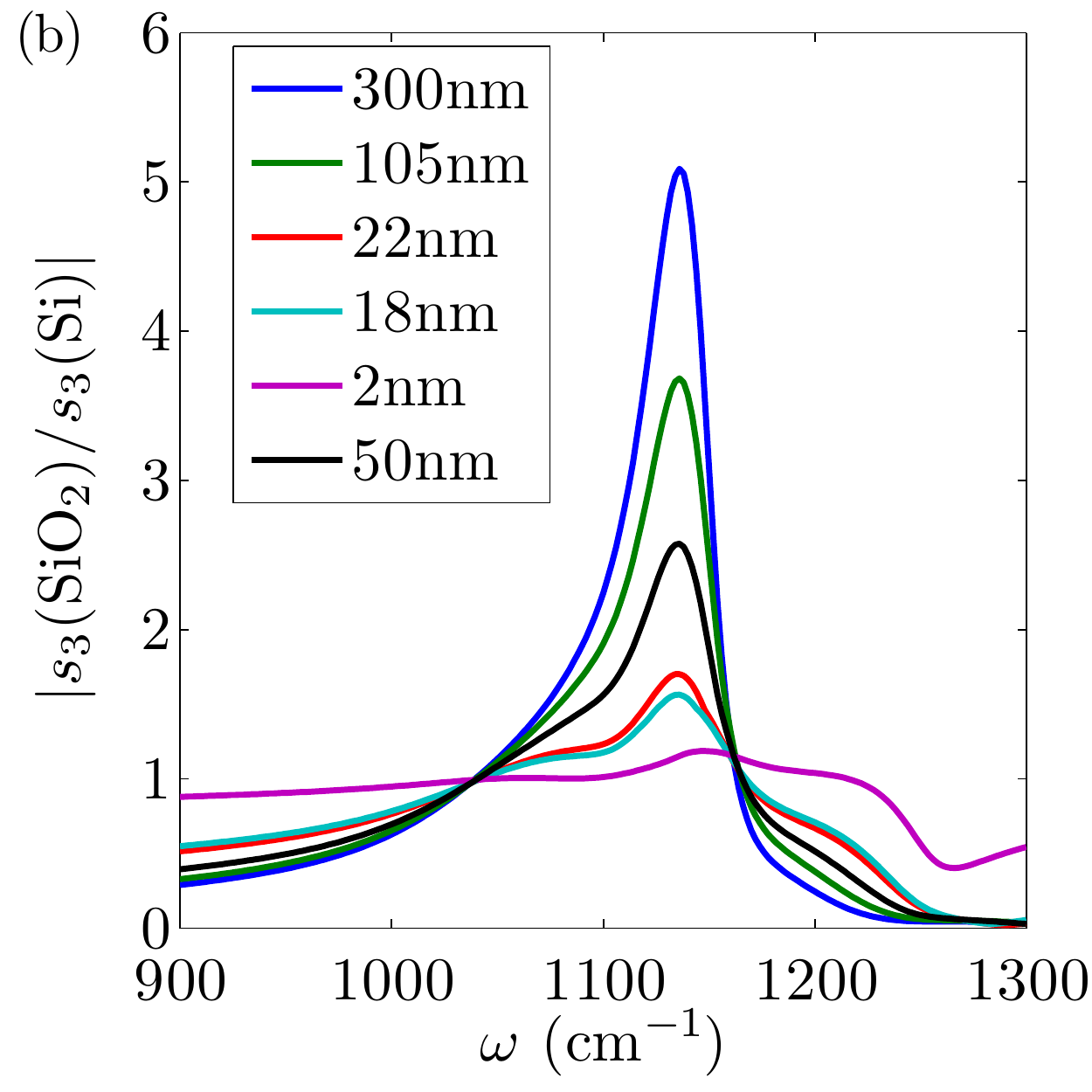}
    \includegraphics[width=2.3in]{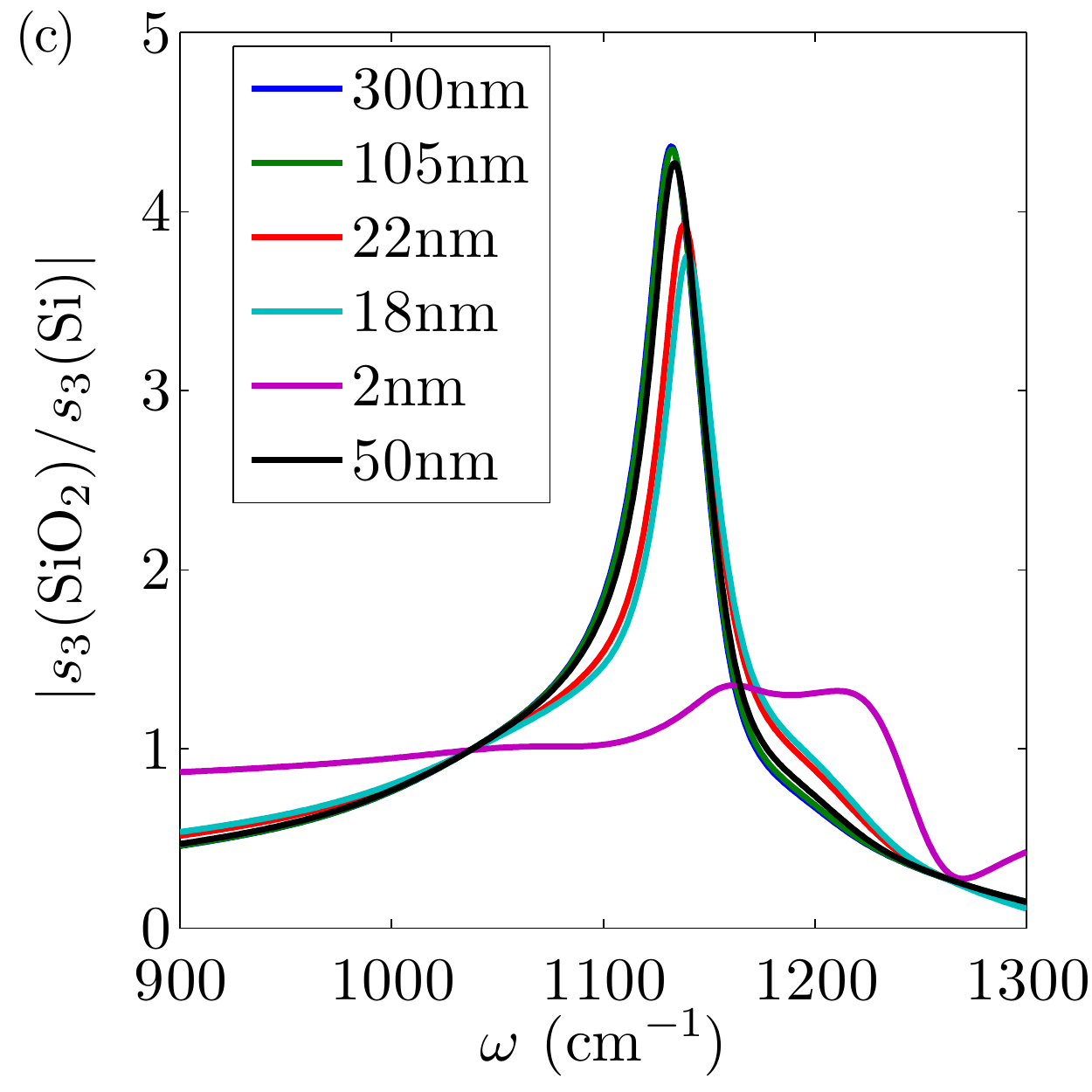}
  %\end{center}
  \caption{(Color online) (a) Main panel: Measured infrared near-field spectra for several \SiO2 film thicknesses.
  The quantity plotted is the absolute value $s_3$ of the third harmonic of the scattering amplitude normalized by that for the Si wafer.
% Inset a: a topographic AFM image of one of the samples. Insets b and c: the near-field images of the same region acquired at the two frequencies
% indicated by the single-headed arrows.
  (b) Theoretical results for the spheroid model with $a = 30\,\text{nm}$ and $L = 15 a$.
  (c) Theoretical results for the point-dipole model with  $a = 30\,\text{nm}$ and $b = 0.75 a$.}
  \label{fig:SpecSOd}
\end{figure*}
%%%%%%%%%%%%%%%%%%%%%%%%%%%%%%%%%%%%%%%%%%%%%%%%%%%%%%%%%%%%%%%%%%%%%%%%%

To make the calculations tractable, we follow examples in the literature~\cite{Porto2003rse, Renger2005rls, Esteban2009fso}
and model the tip as a metallic spheroid of total length $2 L \gg a$, see Fig.~\ref{fig:sSNOM}. As shown below, this 
gives results in a much better agreement with the experiment
in terms of both the frequency and the thickness dependence of the near-field signal. We attribute the origin of the more gradual film-thickness dependence in the spheroidal model to the aforementioned ``antenna effect.'' The magnitude of this effect is determined by the material response over length scales ranging from $a$ to $2 L$, and so it truly saturates only when the film thickness becomes much larger than $2 L$. 

The remainder of the paper is organized as follows.
In Sec.~\ref{sec:Experiment} we summarize the experimental procedures and results.
In Secs.~\ref{sec:Response} and
\ref{sec:Model} we discuss the two theoretical models and compare their predictions with the measurements. Concluding remarks are given in Sec.~\ref{sec:Conclusions}.

%%%%%%%%%%%%%%%%%%%%%%%%%%%%%%%%%%%%%%%%%%%%%%%%%%%%%%%%%%%%%%%%%%%%%%%%%%%%%
\section{Experiment}
\label{sec:Experiment}

To make the paper self-contained we summarize the results of our recent experiments~\cite{Andreev2011xxx} in this Section. We investigated commercially available calibration
 gratings, which contain strips or islands of \SiO2 thermally grown on Si.
% , see the inset in Fig.~\ref{fig:SpecSOd}(a).
The  manufacturer specified thicknesses of the \SiO2 layer spanned the range $d_1 = 2$, $18$, $22$, $108$, and $300\,\text{nm}$.
A combination of CO$_2$ and tunable quantum cascade lasers (\textsc{Daylight Solutions}) allowed us to cover the frequency range between
890 $\icm$ and 1250 $\icm$. The near-field data were collected using a \textsc{Neaspec} system.

\begin{figure*}
  \begin{center}
    \includegraphics[width=2.3in]{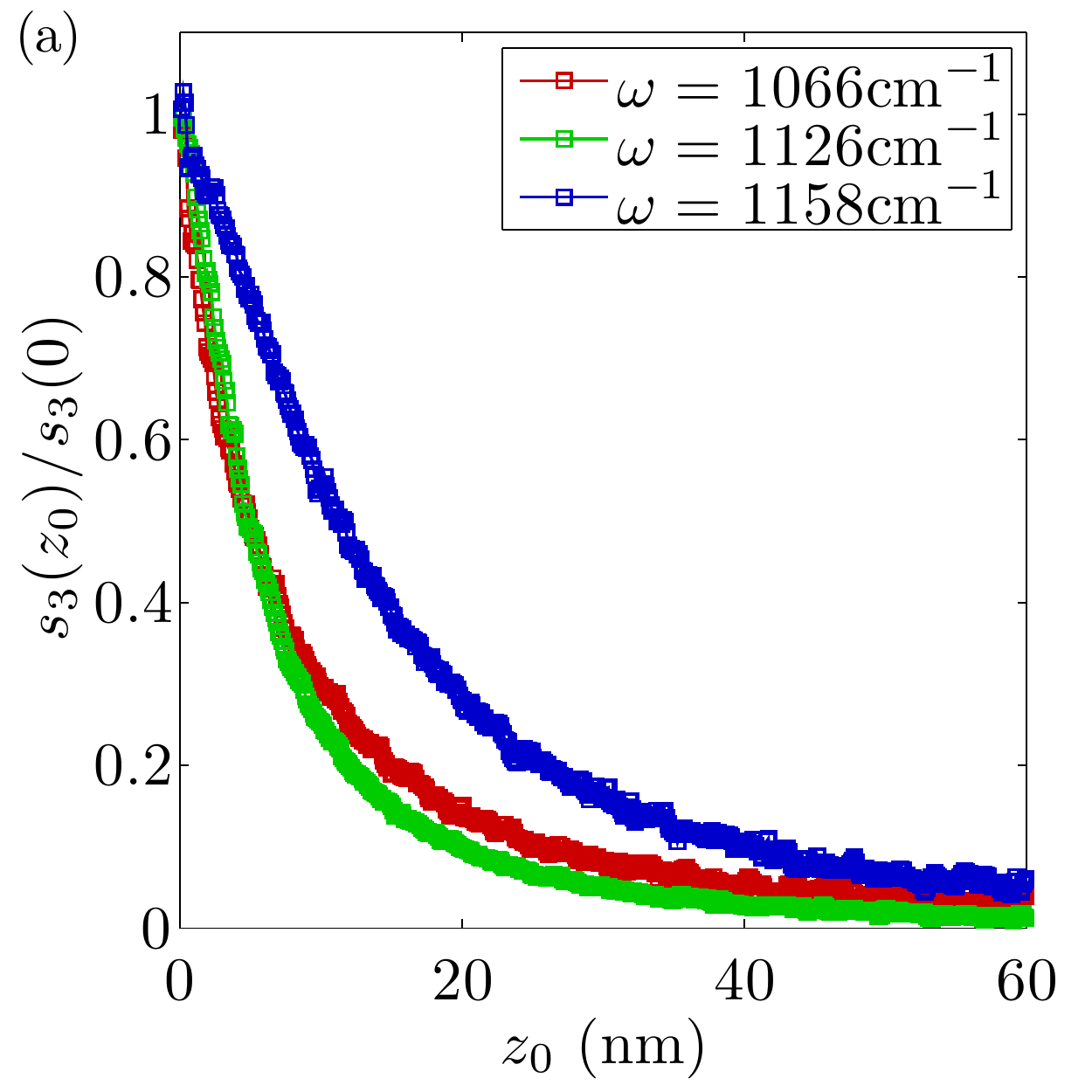}
    \includegraphics[width=2.3in]{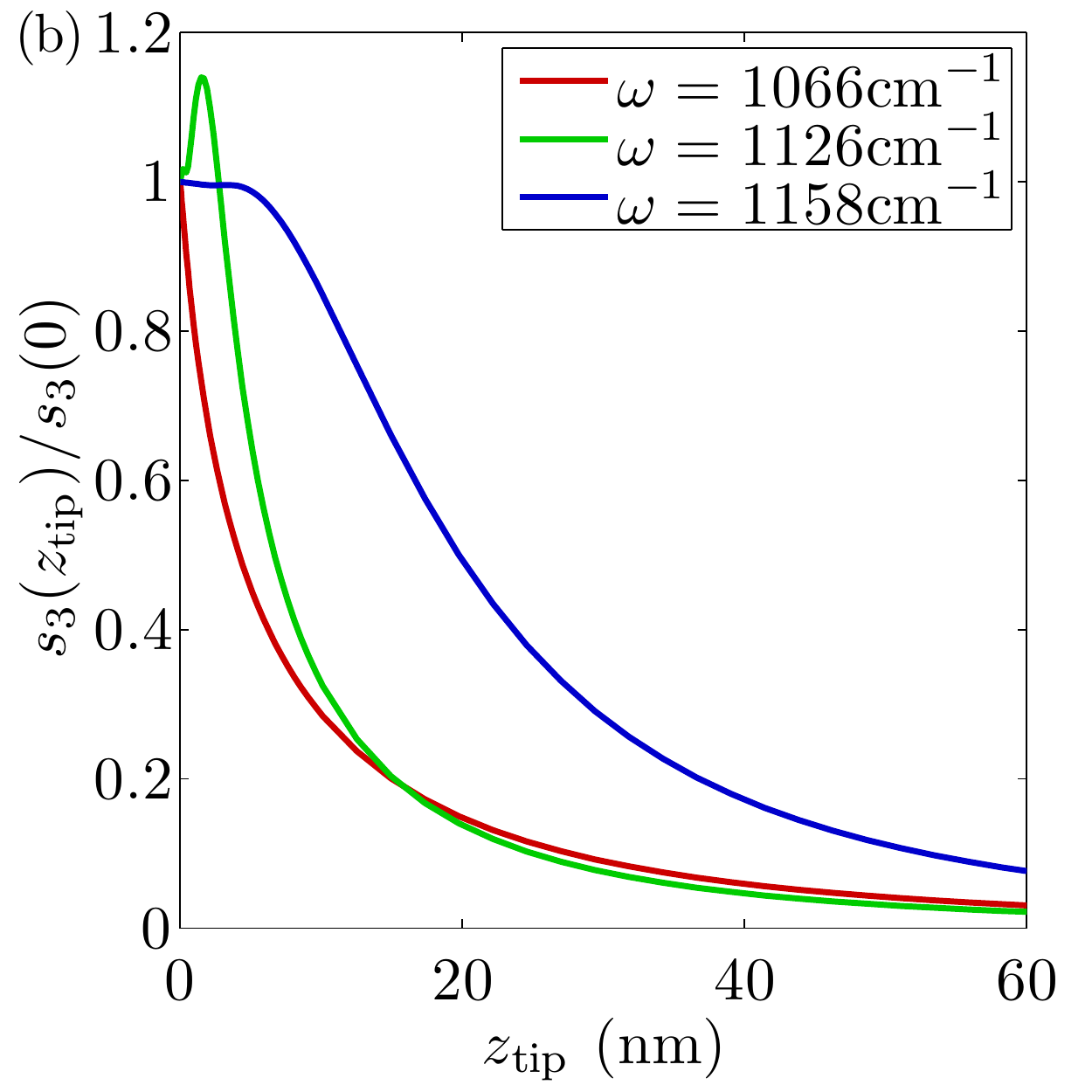}
    \includegraphics[width=2.3in]{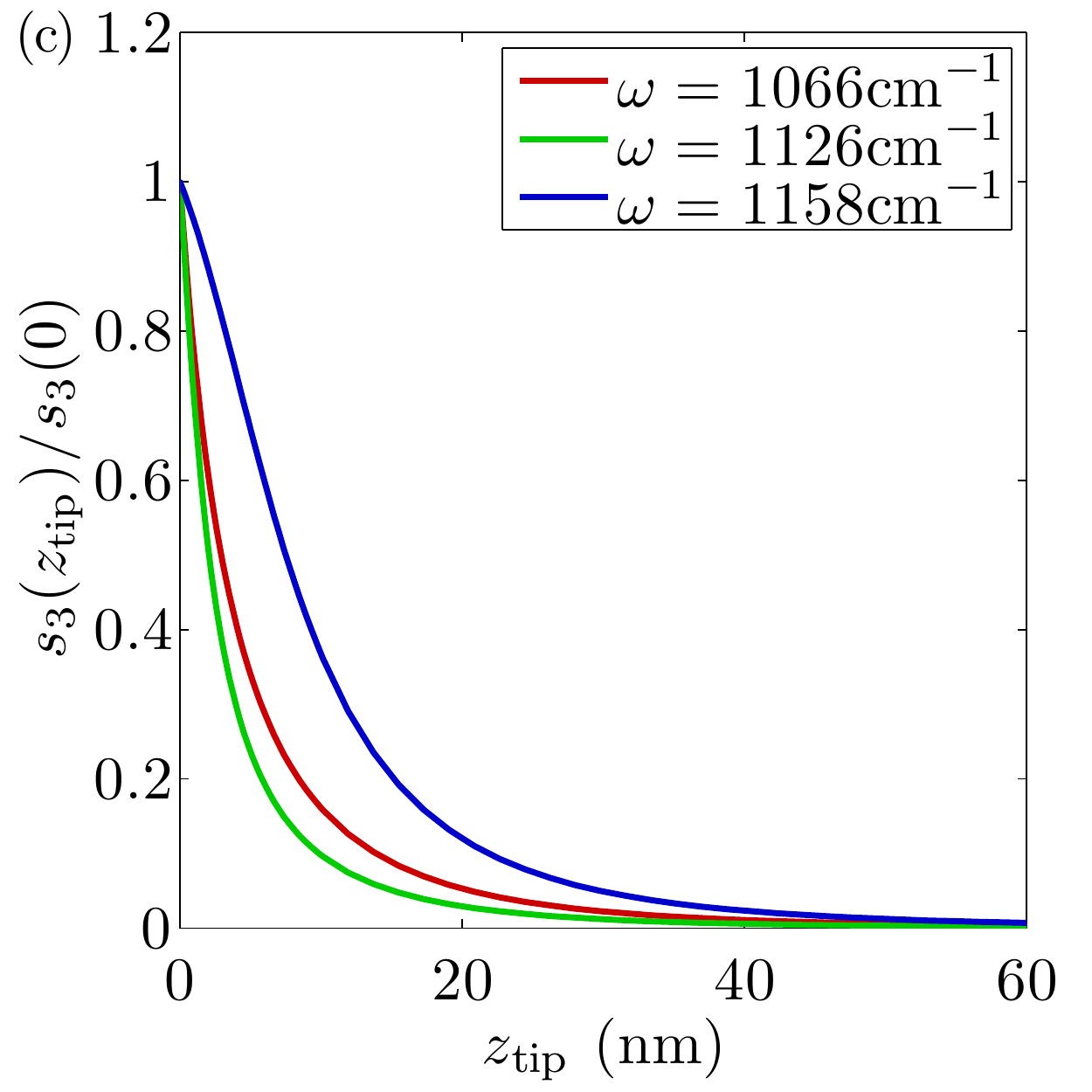}
  \end{center}
  \caption{(Color online) Approach curves. (a) Experimental data for the 105-nm thick \SiO2.
  Theoretical results for the spheroid (b) and the point-dipole models (c) using the same parameters as in Fig.~\ref{fig:SpecSOd}.}
  \label{fig:ApchSOd}
\end{figure*}

The measured s-SNOM signal represents the electromagnetic field backscattered by the probe and the scanned sample. The complex amplitude $s(\omega, t)$ of the backscattered field varies periodically with the tapping frequency $\Omega \sim 40\,\text{kHz}$ as the distance $z_{\text{tip}}$ between the sample and the nearest point of the tip undergoes harmonic oscillations
\begin{equation}
z_{\text{tip}}(t) = z_0 + \Delta z\, (1 - \cos \Omega t)\,,
\label{eqn:z}
\end{equation}
where $\Delta z = 50\,\text{nm}$ typically. In order to suppress unwanted background and isolate the part of the signal scattered by the probe tip, the signal is demodulated. Namely, we extracted
the absolute values $s_n(\omega)$ and phases $\phi_n(\omega)$ at tapping harmonics
\begin{equation}
s_n e^{i \phi_n} = \int\limits_0^{T} \frac{d t}{T}\, e^{i n \Omega t}\, s(\omega, t)\,,
\quad
T = \frac{2\pi}{\Omega}\,.
\label{eqn:s_n}
\end{equation}
The experimental results for the spectra are shown in Fig.~\ref{fig:SpecSOd}(a).
These spectra were intended to be taken in the tapping mode, i.e., for zero $z_0$. However, experimentally $z_0$ can be determined only up to an additive constant $\sim 1\,\text{nm}$. Therefore, we measured $z_0$-dependence of $s_3$ (the approach curves) shown in Fig.~\ref{fig:ApchSOd}(a) and selected the largest observed $s_3$.
Our results are in a qualitative agreement with previous experimental study,~\cite{Taubner2005nrs} which reported approach curves for \SiO2 at a few discrete frequencies and film thicknesses. 

%%%%%%%%%%%%%%%%%%%%%%%%%%%%%%%%%%%%%%%%%%%%%%%%%%%%%%%%%%%%%%%%%%%%%%%%%%
% FIG. 3
%
 %%%%%%%%%%%%%%%%%%%%%%%%%%%%%%%%%%%%%%%%%%%%%%%%%%%%%%%%%%%%%%%%%%%%%%%%%%%%%

% Each spectral data point in Fig.~\ref{fig:SpecSOd}(a) was extracted
% from s-SNOM images that were acquired at a single frequency of a
% tunable infrared laser. Two representative s-SNOM images are shown in
% insets b and c of Fig.~\ref{fig:SpecSOd}(a). The imaged area was
% chosen such that it contained areas of Si and \SiO2 as identified from
% the simultaneously acquired AFM topography images, see inset a of
% Fig.~\ref{fig:SpecSOd}(a).

The data points in Fig.~\ref{fig:SpecSOd}(a) represent the normalized amplitude $s_3(\text{SiO}_2) / s_3 (\text{Si})$, where $s_{3} (\text{SiO}_2)$ and $s_{3} (\text{Si})$ are the raw third-order demodulation signals averaged over the entire \SiO2 and Si areas, respectively. The statistical uncertainty of these averaged data traces is about $2\%$.

For each thickness studied, the normalized amplitude $s_{3} (\text{SiO}_2) / s_3 (\text{Si})$ exhibits several maxima. The main peak is situated at $\omega \approx 1130\,\icm$. The key aspect of the data is a rapid
decrease in the normalized amplitude of this peak as the thickness is
reduced. A trace of this resonance can be reliably identified even for the $2$-$\text{nm}$ thick \SiO2 film. Another notable feature is the growing strength and frequency shift of the secondary peaks on the high-$\omega$ side of the main peak as $d_1$ is decreased.

Since the response of Si is frequency independent in our experimental range, the frequency dependence of the spectra in Fig.~\ref{fig:SpecSOd}(a) originates from that of \SiO2. We attribute the maxima of $s_3 (\text{SiO}_2) / s_3 (\text{Si})$ to the phonon modes localized at the air-\SiO2 interface.~\cite{Amarie2011bia} These resonances occur in the frequency region between the bulk transverse and longitudinal modes of \SiO2 (the outer dashed lines in Fig.~\ref{fig:Im_r_P} below).

The results of our theoretical calculations for the normalized scattering amplitude are presented in the remaining panels of Figs.~\ref{fig:SpecSOd} and \ref{fig:ApchSOd}. They are discussed and compared with the experimental findings in the following Sections.

%%%%%%%%%%%%%%%%%%%%%%%%%%%%%%%%%%%%%%%%%%%%%%%%%%%%%%%%%%%%%%%%%%%%%%%%%%%%%
\section{Response functions and collective modes}
\label{sec:Response}

The sample is modeled as a two-layer system.
The first layer with dielectric function $\epsilon_1(\omega)$ occupies the slab $-d_1 < z < 0$.
The second layer with dielectric function $\epsilon_2(\omega)$ occupies the half-space $z < -d_1$.
The half-space $z > 0$ (``layer 0'') is filled with air (dielectric constant $\epsilon_0 = 1$). The fundamental response functions of the system
are the reflection coefficients $r_X(q, \omega)$, which are functions of in-plane momentum $q$, frequency $\omega$, and polarization $X = S$ or $P$.
The domain of definition of $r_X(q, \omega)$ is understood to include nonradiative modes $q > \sqrt{\epsilon_0}\,\omega / c$. It is known from previous studies that the s-SNOM signal is dominated by the $P$-polarized waves. In our two-layer model their reflection coefficient is given by a Fresnel-like formula
\begin{align}
  r\tP(q,\omega) &= \frac{\eps* \kz0 - \eps0 \kz1}{\eps* \kz0 + \eps0 \kz1} \,,
\label{eqn:r_P_full}\\
  \eps*(q,\omega) &= \eps1 \,
  \frac{\eps2 \kz1 - \eps1 \kz1 \tanh i\kz1 d_1 }
  {\eps1 \kz2 - \eps2 \kz1 \tanh i \kz1 d_1} \,,
\label{eqn:eps_eff}
\end{align}
where $z$-axis momenta $\kz{j}$ are defined by
\begin{equation}
  \kz{j} = \sqrt{\eps{j}\, \frac{\omega^2}{c^2} - q^2}\,,
\quad
\text{Im}\, \kz{j} \geq 0\,.
\label{eqn:k_j}
\end{equation}
Equation~\eqref{eqn:r_P_full} is valid for arbitrary $q$. In the near-field case where $q$ is large and $\kz{j} \simeq i q$, it simplifies to
\begin{equation}
  r\tP(q,\omega) \simeq \frac{\eps* - \eps0}{\eps* + \eps0}
  \,,\quad
  \eps* \simeq \eps1
  \frac{\eps2 + \eps1 \tanh q d_1}{\eps1 + \eps2 \tanh q d_1}\,.
\label{eqn:r_P_near}
\end{equation}
Assuming all $\epsilon_j$ are $q$-independent, the effective dielectric function $\eps*(q,\omega)$ depends on $q$ only via the product $q d_1$ in this limit. 
Therefore, $r\tP(q, \omega)$ for one thickness $d_1$ can be obtained from another by rescaling $q$. As discussed in Sec.~\ref{sec:Introduction} and shown in more detail below, the most important momenta are $q \sim 1 / a$ where $a \sim 30\,\text{nm}$ is the tip radius. Therefore, we can get an approximate understanding of the system response by examining the behavior of $r\tP(q, \omega)$ as a function of $\omega$ at fixed $q d_1 \sim d_1 / a$. This behavior is dictated by the spectrum of surface collective modes, as follows.

In general, surface modes correspond to poles of the response functions $r_X$. Function $r\tP$ given by Eq.~\eqref{eqn:r_P_near} can have up to two poles at each $q d_1$, see, e.g., Ref.~\onlinecite{Prade1991gow}. They are defined by the following condition on $\eps1(\omega)$:
\begin{equation}
\eps1(\omega) = -\frac{\eps0 + \eps2}{2\tanh q d_1}
                 \pm \sqrt{\frac{(\eps0 + \eps2)^2}{4\tanh^2 q d_1} - \eps0 \eps2}
                 \,.
\label{eqn:SP_condition}
\end{equation}
At large $q d_1$, where $\tanh q d_1 = 1$, this condition yields $\eps1(\omega) = -\eps0$ or $\eps1(\omega) = -\eps2$, which correspond to modes localized at the upper $0$--$1$ and the lower $1$--$2$ interfaces, respectively. Actually, the latter ``pole'' has vanishingly small residue because evanescent waves do not reach the lower interface at $q d_1 = \infty$.
There is no $q$-dispersion and no coupling of the two modes in this limit. The dispersion appears at finite $q d_1$, where the two modes become mixed. In particular, we find
\begin{subequations}
\label{eqn:SP_small_q}
\begin{align}
\eps1(\omega) &\simeq -\frac{q d_1}{\eps0^{-1} + \eps2^{-1}}\,,
&& \text{``0--1''}
\label{eqn:SP_small_q_01} \\ 
  &\simeq -\frac{\eps0 + \eps2}{q d_1}
&& \text{``1--2''}
\label{eqn:SP_small_q_12}
\end{align}
\end{subequations}
at $q d_1 \ll 1$. At finite $q$, both interfaces participate in generating these excitations. The labels ``0--1'' and ``1--2'' are for convenience: they indicate at which interface a given dispersion branch is ultimately localized as $q$ increases. At $q d_1 = 0$, the ``0--1'' and ``1--2'' branches are characterized by $\eps1(\omega) = 0$ and $\eps1(\omega) = -\infty$, which correspond, respectively, to the bulk longitudinal and transverse phonon frequencies $\omega\tsub{LO}$ and $\omega\tsub{TO}$. 

If we try to apply this formalism to real materials, we face the problem that Eq.~\eqref{eqn:SP_condition} has no solutions for real $\omega$ because the dielectric functions have finite imaginary parts. This is why in practice the collective mode spectra are usually defined differently. They are identified with the maxima of dissipation, i.e., $\text{Im}\, r\tP$. The number of these maxima can be fewer than the total allowed number of the modes because some of them can be overdamped. Similarly, we define $\omega\tsub{LO}$ and $\omega\tsub{TO}$ as the frequencies that correspond to the maxima of $-\text{Im}\,\eps1^{-1}(\omega)$ and $\text{Im}\,\eps1(\omega)$.

\begin{figure}[b]
  \begin{center}
    \includegraphics[width=2.8in]{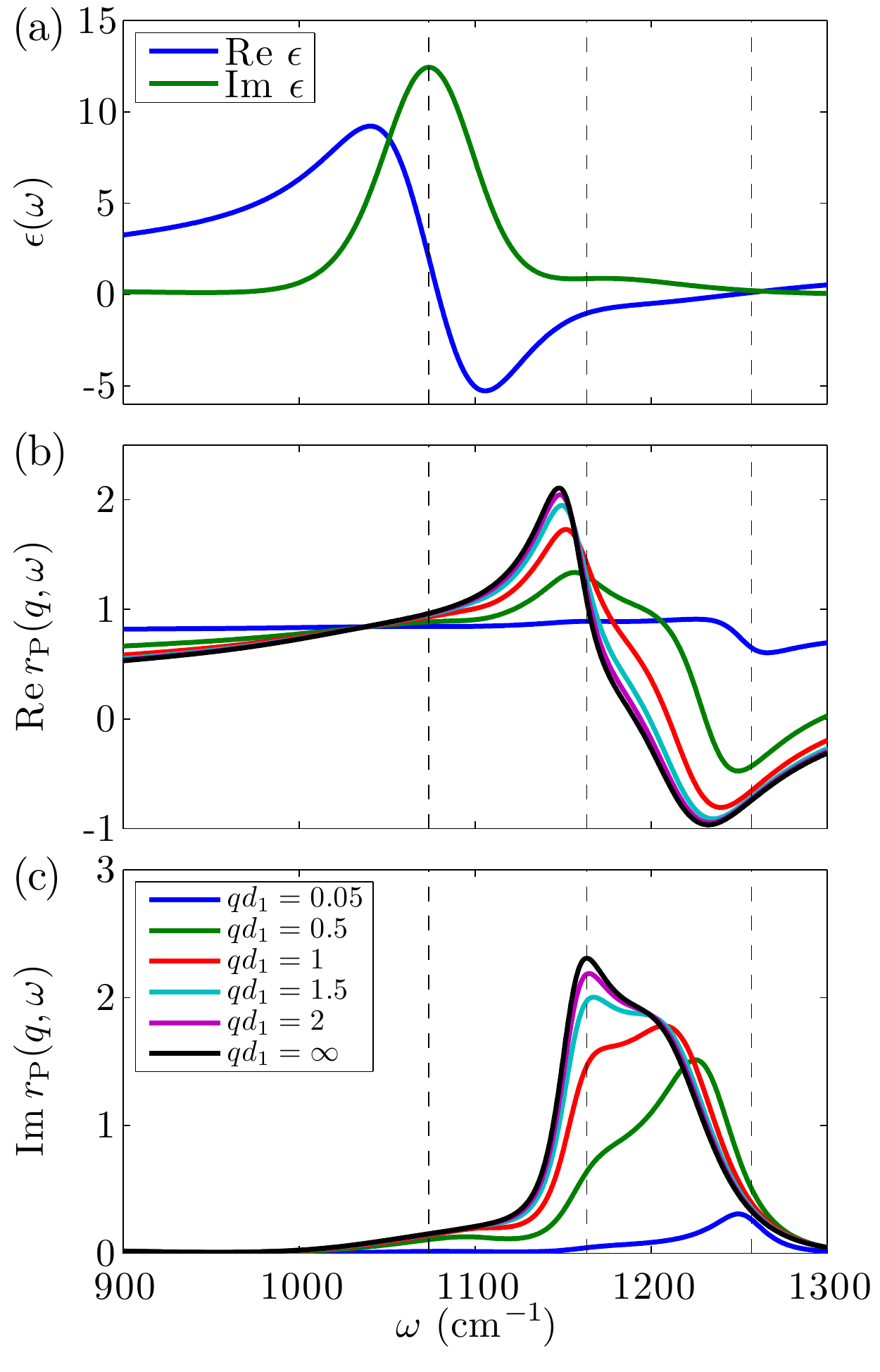}
  \end{center}
  \caption{(Color online) (a) Dielectric function of bulk \SiO2\ as a function of frequency from ellipsometry. (b) The real and (c) the imaginary parts of the near-field reflection coefficient, $r\tP(q, \omega)$ for several $q d_1$.  In all the panels three dashed lines indicate the transverse optical phonon frequency $\omega\tsub{TO} \approx 1074\,\icm$, the surface optical phonon frequency $\omega\tsub{SP} \approx 1164\,\icm$, and the longitudinal optical phonon frequency $\omega\tsub{LO} \approx 1263\,\icm$.
 \label{fig:Im_r_P}}
\end{figure}

To see what kind of spectra are realized in our system, we use
our ellipsometry data for $\eps1(\omega)$ [Fig.~\ref{fig:Im_r_P}(a)] and Eq.~\eqref{eqn:r_P_near} to compute $r\tP$ for several values of $q d_1$. The plot of these quantities as a function of $\omega$ is presented in Fig.~\ref{fig:Im_r_P}(c). Three maxima on each curve in the region of primary interest $\omega > 1000\,\icm$ are apparent. They exist already at $q d_1 = \infty$, and so all of them belong to the upper (air-\SiO2) interface. In fact, we do not expect sharp modes at  the lower (\SiO2-Si) interface because the dielectric function of Si is quite large $\eps2 \approx 11.7$ in the studied range of $\omega$. The lowest value of $\text{Re}\,\eps1 \approx -5.0$ is not sufficient to compensate $\eps2$ and generate ``1-2'' modes, cf.~Eq.~\eqref{eqn:SP_condition}.

%%%%%%%%%%%%%%%%%%%%%%%%%%%%%%%%%%%%%%%%%%%%%%%%%%%%%%%%%%%%%%%%%%%%%%%%%%
% FIG. 4
%
%%
%%%%%%%%%%%%%%%%%%%%%%%%%%%%%%%%%%%%%%%%%%%%%%%%%%%%%%%%%%%%%%%%%%%%%%%%%

The main peak of $\text{Im}\, r\tP$ at $q d_1 = \infty$ defines the surface phonon frequency of \SiO2 $\omega\tsub{SP} \approx 1164\,\icm$.
There also exist secondary peaks at $\omega \approx 1100\,\icm$ and $\omega \approx 1220\,\icm$. Their evolution as a function of $q d_1$ comply with the general scheme outlined above. As $q d_1$ decreases, all the three peaks loose strength, as expected, because the amount of \SiO2 diminishes. The lower-$\omega$ secondary peak redshifts, moving towards $\omega\tsub{TO}$, and then quickly disappears. This agrees with the \SiO2-Si resonance being highly damped. The higher-$\omega$ secondary peak becomes dominant at $q d_1 < 0.5$ and demonstrates a systematic shift towards $\omega\tsub{LO}$, see Fig.~\ref{fig:Im_r_P}(c).

A notable feature of Fig.~\ref{fig:Im_r_P}(b) is the clustering of the crossing points of the different curves near $\omega = 1036\,\icm$. This is the frequency where the dielectric function of \SiO2 is the closest to that of Si, $\eps2 \approx 11.7$. As a result, the two layers act almost as one bulk material, so that $r\tP(\omega)$ is approximately thickness-independent.

There is a qualitative correspondence between the features displayed by the reflection coefficient $r\tP$ and the observed near-field signal $s_3 (\text{SiO}_2) / s_3 (\text{Si})$, cf.~Figs.~\ref{fig:SpecSOd}(a) and \ref{fig:Im_r_P}(b),(c). However, the relation between $r\tP(q, \omega)$ and the measured s-SNOM signal is nontrivial. For example, the frequency positions of the maxima in $\text{Im}\,r\tP(q, \omega)$ and those in $s_3 (\text{SiO}_2)/ s_3 (\text{Si})$ differ by as much as $40\,\icm$. We also suspect that there may be some slight differences between the optical constants of thick films we assume in our calculations and those of the small \SiO2 structures we probe by the s-SNOM. This is the likely reason why the crossing point of the experimental curves occurs near $1060\,\icm$ rather than $1036\,\icm$ predicted by both our models, cf.~Fig.~\ref{fig:SpecSOd}.

Developing a reliable procedure for inferring $r\tP(q, \omega)$ from $s_3$ remains a challenge for the theory. The next section presents our current approach towards this ultimate goal.

\section{Tip-sample interaction}
\label{sec:Model}

Both radiative and nonradiative waves may play significant roles in the s-SNOM experiment.~\cite{Porto2003rse} The radiative modes magnify the signal by a certain far-field factor (FFF) $F(q_s, \omega)$, where
$q_s = (\omega / c) \sin \theta$ is the momentum of these modes for the angle of incidence $\theta$. The nonradiative modes influence the effective polarizability $\chi(\omega, z_{\text{tip}})$ of the tip, i.e., the ratio of its dipole moment $p^z$ and the external electric field $E_{\text{ext}}^z$. Altogether the demodulated s-SNOM signal $s_n e^{i \phi_n}$ can be written as
\begin{align}
s_n e^{i \phi_n} &\propto \chi_n E_{\tExt} \sin 2 \theta\,
                           F(q_s, \omega)\,,
\label{eqn:S_from_chi}\\
{\chi}_n(\omega) &= \int\limits_0^{T} \frac{d t}{T}\, e^{i n \Omega t}\, {\chi}\bigr(\omega, z_{\text{tip}}(t)\bigl)\,.
\label{eqn:chi_n}
\end{align}
Below we discuss the FFF and the tip polarizability separately.

\subsection{Far-field factor}
\label{sec:Far}

\begin{figure}[t]
  \begin{center}
    \includegraphics[width=2.6in]{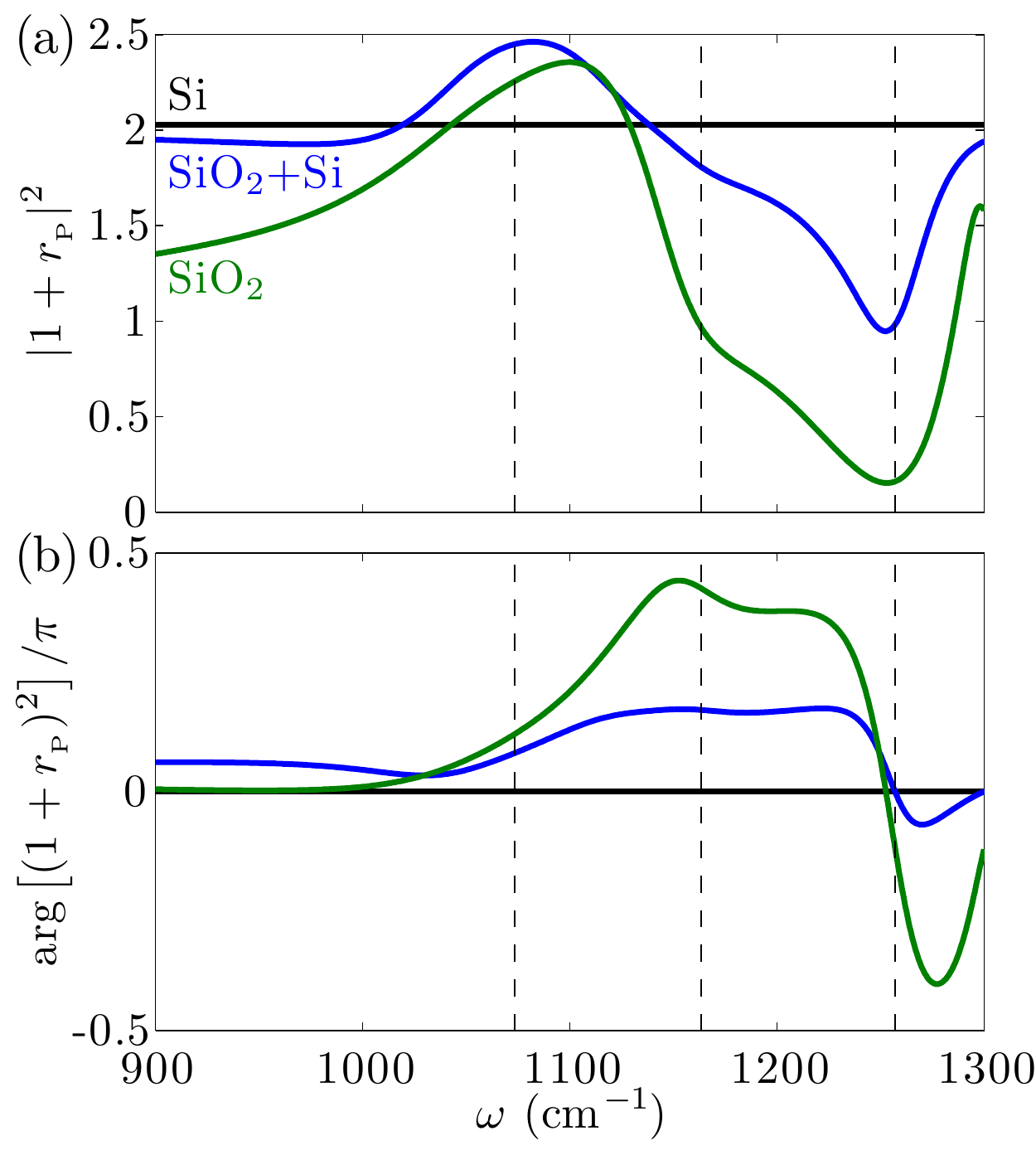}
  \end{center}
  \caption{(Color online) (a) The absolute value and (b) the phase of the far-field factor computed as a function of frequency for the incidence angle $\theta = 45^\circ$.
The black trace is for bulk Si substrate, the blue one is for bulk \SiO2 substrate, the green one is for $300\,\text{nm}$ thick \SiO2 followed by bulk Si. The meaning of the dashed lines is the same as in Fig.~\ref{fig:Im_r_P}.
 \label{fig:r_P}}
\end{figure}

The FFF for an infinite layered system is given by~\cite{Sukhov2004rom, Aizpurua2008sei}
\begin{equation}
F(q_s, \omega) = [1 + r\tP(q_s, \omega)]^2\,.
\label{eqn:F}
\end{equation}
%%
%
% Since layer 1 is only a thin film of \SiO2, the FFF is dominated by the
% layer 2, i.e., Si, which has a frequency-independent response in the
% frequency range studied. Thus, the total FFF is also fairly constant.
% Nevertheless, as known from ellipsometry, even a thin film can cause
% detectable variations of FFF when the frequency passes through its
% resonance.
As shown in Figs.~\ref{fig:r_P}(a), for $d_1 = 300\,\text{nm}$ \SiO2 film, the absolute value of the FFF has a maximum near $\omega_{\text{\,TO}} \approx 1074\,\icm$ and a suppression near $\omega_{\text{\,LO}} \approx 1272\,\icm$. For thinner films, these features are less pronounced.
The main maximum of $s_3$, which is the main focus of our analysis, is away from both $\omega_{\text{\,TO}}$ and $\omega_{\text{\,LO}}$. It is essentially unaffected by the FFF. Still, if FFF were to be included in the calculation in the form prescribed by Eq.~\eqref{eqn:F}, it would produce a visible hump of $s_3(\omega)$ near $\omega_{\text{\,TO}}$ and a dip near $\omega_{\text{\,LO}}$. These features are not present in the experimental data, Fig.~\ref{fig:SpecSOd}(a). A better agreement with the experiment is obtained if $F(q_s, \omega)$ is set to a constant, which is what we do here. We rationalize this decision by noting that the \SiO2 layer in the actual samples does not extend over the entire $x$--$y$ plane but occupies only small sub-wavelength regions.
%, see inset a of Fig.~\ref{fig:SpecSOd}(a).
Therefore, the FFF is dominated by the $\omega$-independent response of Si.

%%%%%%%%%%%%%%%%%%%%%%%%%%%%%%%%%%%%%%%%%%%%%%%%%%%%%%%%%%%%%%%%%%%%%%%%%%
% FIG. 5
%
%%
%%%%%%%%%%%%%%%%%%%%%%%%%%%%%%%%%%%%%%%%%%%%%%%%%%%%%%%%%%%%%%%%%%%%%%%%%%%%%

\subsection{Point-dipole model of the tip}
\label{sec:Point}

The effective tip polarizability $\chi(\omega, z_{\text{tip}})$ is the most important factor on the right-hand side of Eq.~\eqref{eqn:S_from_chi} and it is also the most difficult one to compute. This quantity is dictated by the near-field coupling between the tip and the sample. For irregular tip shapes it can be calculated only numerically. However, previous s-SNOM studies demonstrated that acceptable results can often be obtained if the tip is approximated by a spheroid,~\cite{Porto2003rse, Renger2005rls, Esteban2009fso} a small sphere,~\cite{Rendell1981spc, Aravind1982uop, Aravind1983teo, Ruppin1992oab, Sukhov2004rom, Renger2005rls} a ``finite'' dipole,~\cite{Cvitkovic2007amf, Amarie2011bia} or a point dipole.~\cite{Hillenbrand2000coc, Taubner2004nrt, Aizpurua2008sei} The actual tip shape in our experiment is close to a rounded pyramid.

The point-dipole approximation is the simplest one and it has been used extensively for modeling s-SNOM experiments, including those performed on multilayer systems.~\cite{Aizpurua2008sei, Fei2011ino} The point-dipole model has two adjustable parameters: the polarizability $a^3$ of the effective dipole and its position $b$ with respect to the bottom of the tip. The results obtained following the standard analysis~\cite{Aizpurua2008sei, Fei2011ino} are shown in Fig.~\ref{fig:SpecSOd}(c) using $a = 30\,\text{nm}$ and $b = 0.75 a$.
We see that even for this rather large $a$ the point-dipole model does not reproduce the observed strong dependence of $s_3$ on thickness at $d_1 > 22\,\text{nm}$.

The discrepancy can be seen more clearly in Fig.~\ref{fig:SOd_maxS3}, where the height of the peak in $s_3(\text{SiO}_2) / s_3 (\text{Si})$ corresponding to the surface phonon is plotted as a function of $d_1$. For the point dipole model the curve flattens at $d_1 \sim b$. In contrast, the experimentally observed $s_3 (\text{SiO}_2)/ s_3 (\text{Si})$ maximum continues to rise with $d_1$. The point-dipole model also predicts a very steep approach curve, Fig.~\ref{fig:ApchSOd}(c), in poor agreement with the measurements.

The physical origin of the saturation of the thickness dependence in Fig.~\ref{fig:SpecSOd}(c) is easy to understand. One can think about the near-field coupling between the point dipole and the sample in terms of the method of images. For a dipole positioned at $z_{\text{pd}} = z_{\text{tip}} + b$, the image is concentrated at the depth $z_{\text{pd}}$ below the surface. Therefore, films of thickness larger than $z_{\text{pd}}$ would act as a bulk material. Another way to arrive at the same conclusion is to notice that the characteristic range of momenta of the relevant nonradiative waves is $q \lesssim 1 / z_{\text{pd}}$.
Since $r\tP$ depends on $q$ through the term $\tanh q d_1$ [Eq.~\eqref{eqn:r_P_near}], the dependence of the near-field coupling on $d_1$ should saturate at $d_1 \gtrsim z_{\text{pd}} \sim b$.

\subsection{Spheroid model of the tip}
\label{sec:Spheroid}

The lack of saturation in the observed s-SNOM signal as a function of $d_1$ at $d_1 \gg a$ indicates that evanescent waves with momenta $q \ll 1 / a$ also play an important role in the near-field coupling between the tip and the sample. This is a signature of models in which the tip has a finite extent in space $2 L \gg a$, see Fig.~\ref{fig:sSNOM}. Although such models are certainly more realistic than a point-dipole approximation, there has not been a systematic study of how the results would depend on the exact shape of the tip. Given some initial success of the point-dipole approximation, we speculate that a suitable simple shape can provide a good compromise between increase in computational effort and ability to capture relevant physics.

To test this idea, we model the tip as an elongated metallic spheroid positioned above a two-layer medium. This follows a tradition in the literature wherein similar models were considered~\cite{Porto2003rse, Renger2005rls, Esteban2009fso} for the case of bulk substrates. In Ref.~\onlinecite{Cvitkovic2007amf} an analytical formula for the spheroidal tip was also proposed, based on heuristic arguments. However, it cannot be easily extended to the $q$-dependent $r\tP$ we study here. Instead, our calculations are done numerically. They involve only two essential approximations. One is neglecting retardation, which is justified is the length $2L$ of the spheroid is smaller than $\lambda$. The other one is neglecting the finite skin depth of the metal (Pt-Ir alloy) covering the tip. Due to computational difficulties involved, this issue is left for future investigation.

The calculations were performed in two ways. First is the standard boundary-element method. In this method we divide the entire tip --- assuming azimuthal symmetry --- into a large number (typically, 200) of small cylindrical segments. We assume that different segments interact by Coulomb interaction as coaxial rings. The interaction of each segment with itself is defined in such a way that the polarizability of the tip in the absence of the sample coincides with the known analytical result for the prolate spheroid.
The effect of the sample is included by adding ring-ring interactions mediated by reflected electrostatic fields. This is accomplished by numerical quadrature over the product of $r\tP(q, \omega)$ and suitable form-factors.
This is the most time-consuming step of the simulation.
After the interaction kernel is generated in this way, it is straightforward to solve numerically for the dipole moment of the tip induced by a unit external field, which is the desired polarizability $\chi(\omega, z_{\text{tip}})$.

We also developed a second numerical method of computing $\chi$ (to be described elsewhere), based on an expansion of the electric field in ellipsoidal harmonics. This alternative method is similar to that used for a metallic sphere above a dielectric half-space.~\cite{Ford1984eio, Sukhov2004rom} We verified that the two methods give identical results.

Substituting the computed polarizability $\chi$ into Eqs.~\eqref{eqn:S_from_chi} and demodulating per Eq.~\eqref{eqn:chi_n}, we obtain
approach curves. Figure~\ref{fig:ApchSOd}(b) illustrates that some approach curves are nonmonotonic near the resonances. In calculating the s-SNOM amplitude $s_3$ we choose $z_{\text{tip}}$ that corresponds to the largest $s_3$ because this is how it was done in the experiments. The results for the normalized amplitude are plotted in Fig.~\ref{fig:SpecSOd}(b).
 
\begin{figure}
  \begin{center}
    \includegraphics[width=2.4in]{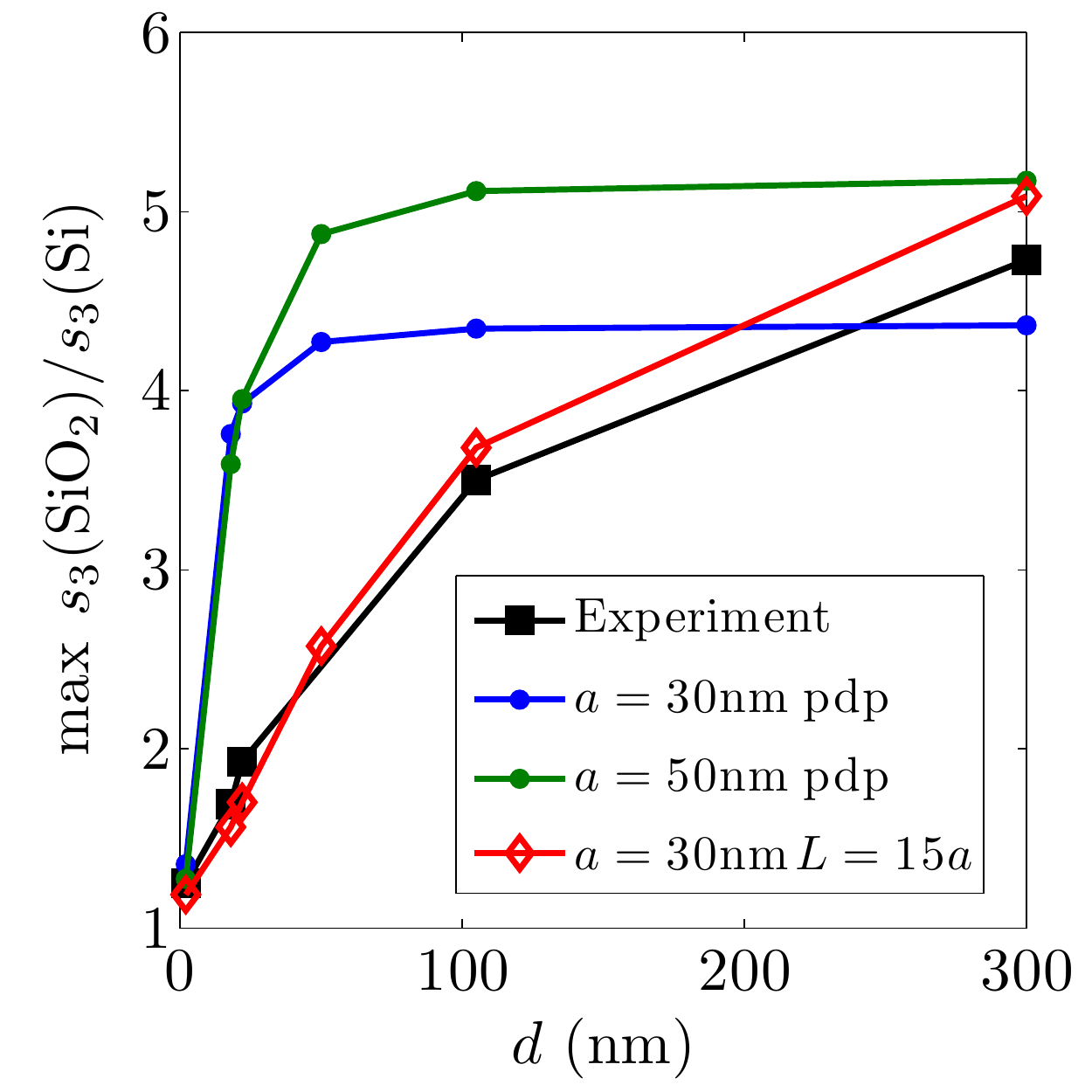}
  \end{center}
  \caption{(Color online) The thickness dependence of the $s_3$ peak for different tip models. The circles represent the point-dipole calculations, one for $a = 30\,\text{nm}$ (blue) and the other for $a = 50\,\text{nm}$ (green). The diamonds are for the spheroid model with  $a = 30\,\text{nm}$ and $L = 15 a$, the same as in Figs.~\ref{fig:SpecSOd}(b). The black squares are derived from the experimental data shown in Fig.~\ref{fig:SpecSOd}(a) after some smoothing over fluctuations.}
  \label{fig:SOd_maxS3}
\end{figure}

The spheroid model has two adjustable parameters: the apex radius of curvature $a$ and the half-length $L$. When $L = a$ the spheroid becomes a sphere.
In this case the spheroid model gives results similar to the point-dipole model, i.e., Fig.~\ref{fig:SpecSOd}(c). As the ratio $L / a$ increases, the differences appear. However, once $L / a$ exceeds ten, the normalized signal $s_3 (\text{SiO}_2) / s_3 (\text{Si})$ does not change much at $d_1 \leq 300\,\text{nm}$. Therefore, for long spheroids we effectively have only a single adjustable parameter, $a$. Remarkably, the thickness dependence of the $s_3$ peak for the spheroid model matches the experiment extremely well (Fig.~\ref{fig:SOd_maxS3}).

%%%%%%%%%%%%%%%%%%%%%%%%%%%%%%%%%%%%%%%%%%%%%%%%%%%%%%%%%%%%%%%%%%%%%%%%%%
% FIG. 6
%
%%
%%
%%%%%%%%%%%%%%%%%%%%%%%%%%%%%%%%%%%%%%%%%%%%%%%%%%%%%%%%%%%%%%%%%%%%%%%%%%%%%

%%%%%%%%%%%%%%%%%%%%%%%%%%%%%%%%%%%%%%%%%%%%%%%%%%%%%%%%%%%%%%%%%%%%%%%%%%%%%
\section{Conclusions}
\label{sec:Conclusions}

In this paper we analyzed the results of experimental study of amorphous \SiO2 films on Si obtained by scanning near-field optical spectroscopy.~\cite{Andreev2011xxx} We discussed the collective mode spectra of such structures and compared measurements with two theoretical calculations.
The first is based on a conventional approximation in which the tip of the scanned probe is modeled as a point dipole. In the second the tip is treated as an elongated spheroid, significantly improving agreement with the experiment.

We explain the qualitative difference between the two models as follows. An important physical ingredient missing in the point-dipole model is the enhancement of the electric field
 near the apex of the tip --- the antenna effect. This phenomenon is well-known from classical electrostatics.
The enhancement of the field is controlled primarily by the ratio of the total length of the tip $2 L$ (actually, the smaller of $2L$ and $\lambda$)
and the apex radius of curvature $\sim a$. The point-dipole model has been successful in the past without this enhancement factor only on account of the
normalization procedure. Instead of absolute $s_n$, one usually reports $s_n$ normalized to some reference material such as Au or in our case, Si. This way, one eliminates
any possible frequency dependence of the source radiation, but at the same time cancels the part of the signal scaling with tip size. For a stratified sample this cancellation
is imperfect because the the field enhancement depends also on the dielectric response of the sample, which is a function of momentum $q$. For a tip of length $2 L$, harmonics
relevant for the field enhancement have momenta ranging from $q \sim 1 / a$ down to $q \sim 1 / L$. Therefore, one may expect that the dependence of the s-SNOM signal on the
thickness $d_1$ of the top layer would saturate only when $d_1 \sim L$. Our simulations provide direct evidence for this claim. Therefore, we think that the spheroid model holds
a great promise as an analysis tool for near-field experiments. It captures a lot of physics relevant to the near-field interaction while remaining computationally fast.

The strong experimentally observed thickness dependence of the near-field signal~\cite{Andreev2011xxx} indicates that s-SNOM is capable of not only high lateral resolution but can also probe the system in the third dimension. However, the response of a layered system is different from those containing small subsurface particles~\cite{Taubner2005nrs}. We hope that experimental and theoretical approaches presented in this paper may be of use for accurate depth profiling of various dielectric and metallic nanostructures.

The work at UCSD is supported by ONR, AFOSR, NASA, and UCOP. AHCN and LMZ acknowledge DOE grant DE-FG02-08ER46512 and ONR grant MURI N00014-09-1-1063. We thank F.~Keilmann and R. Hillenbrand for illuminating discussions.

\bibliography{pubsSNOMSiO2}
\end{document}